\documentclass[aps,prl,tightenlines,onecolumn,groupedaddress,superscriptaddress,showpacs]{revtex4}
\usepackage{amssymb}
\usepackage{color,graphicx}
\usepackage{lmodern}
\usepackage{amsmath}
\usepackage{amsbsy}
\usepackage{amsthm}
\usepackage{float}
\usepackage{bbm}
\usepackage{bm}
\usepackage{epsfig}
\usepackage{caption}

\begin{document}

\title{Double-Slit Interference Pattern for a Macroscopic Quantum System}

\author{Hamid Reza Naeij}
\email[]{naeij@ch.sharif.edu}
\affiliation{Research Group on Foundations of Quantum Theory and Information,
Department of Chemistry, Sharif University of Technology
P.O.Box 11365-9516, Tehran, Iran}
\author{Afshin Shafiee}
\email[Corresponding Author:~]{shafiee@sharif.edu}
\affiliation{Research Group on Foundations of Quantum Theory and Information,
Department of Chemistry, Sharif University of Technology
P.O.Box 11365-9516, Tehran, Iran}

\affiliation{School of Physics, Institute for Research in Fundamental Sciences (IPM), P.O.Box 19395-5531, Tehran, Iran}

\begin{abstract}
In this study, we solve analytically the Schr\"{o}dinger equation for a macroscopic quantum oscillator as a central system coupled to two environmental micro-oscillating particles. Then, the double-slit interference patterns are investigated in two limiting cases, considering the limits of uncertainty in the position probability distribution. Moreover, we analyze the interference patterns based on a recent proposal called stochastic electrodynamics with spin. Our results show that when the quantum character of the macro-system is decreased, the diffraction pattern becomes more similar to a classical one. We also show that, depending on the size of the slits, the predictions of quantum approach could be apparently different with those of the aforementioned stochastic description.

\end{abstract}
\maketitle
\textbf{keywords} Double-slit, Interference, Macroscopic quantum system, Stochastic electrodynamics with spin

\section{Introduction}

Coherency is a property of matter in quantum world which is most demonstrated in a double-slit experiment. This is also important to understand the dual character of quantum matters. Richard Feynman emphasized that the double-slit interference is at the heart of quantum phenomena: ''In reality, it contains the only mystery, the basic peculiarities of all of quantum mechanics'' \cite{Feynman}.

The double-slit diffraction pattern of a micro-particle is a distinguishing point between classical mechanics (CM) and quantum mechanics (QM). The interference fringes of wave-particles are described within QM. This follows from an objective  interpretation of the wave function and by the property that, in QM, superpositions of states are possible, while this is not so in CM \cite{Zecca1, Zecca3}. Interference is resulted from the uncertainty principle between the momentum and the position of the quantum system. If we take the vacuum into account, it will require entanglement between complementary variables such as the momentum and the position or the angular momentum components \cite{Zimmermann}.

Many works have been done to explain double-slit experiment under different conditions such as interference for the macro-system in experimental and theoretical contexts. Nairz, Arndt and Zeilinger studied the interference pattern for the fullerene molecule as a large object similar to a classical system \cite{Nairz}. Hornberger $\it{et}$ $\it{al.}$ investigated the effect of environment on interference  pattern of fullerenes as a macro-molecule \cite{Hornberger2}. Also, Hornberger and others studied quantum interferenece of the clusters in experiment \cite{Hornberger1}. Gerlich and others showed the quantum diffraction of large organic molecules \cite{Gerlich}. 

The double-slit diffraction pattern can be illustrated by describing the incoming particle in some proper states such as Gaussian wave packets \cite {Merzbacher, Holland}. Zecca in many studies investigated theoretically one, two and $N$-slit diffraction patterns for the Gaussian wave packets \cite{Zecca4, Zecca5, Zecca7}. The use of Gaussian states is sufficiently general, because it includes the limit case of plane waves and that of the wave packets narrower than the slit width. Moreover, due to the development of experimental techniques, possible deviations from the standard form of the interference pattern can be better explained by Gaussian-like states \cite{Zecca2}.

In this study, considering a macroscopic quantum oscillator interacting with two micro-oscillating particles in the environment, we exactly obtain  the wave function of the system in the ground state as a Gaussian wave packet. Then, the time evolution for the wave function of the $\it{central}$ system $\it{after}$ the slits is evaluated by considering the $y$-dependence of the wave packet, passing the slits, in two limiting cases. The slits are located on the $y$-axis, in a symmetric position with respect to the $x$-axis. The wave packet describing the incoming particle is factorized in its $x$ and $y$ dependences. It is assumed that $x$ and $y$ dependences of the wave function remain factorized during and after passing the slits \cite{Zecca2}. We show that when quantum character of the macro-system is evanesced, interference fringes are diminished and the diffraction pattern becomes more similar to a classical one.

The double-slit experiment is also a point where one can compare the predictions of QM with those of stochastic electrodynamics with spin (SEDS). The comparison of the two theories was first studied in \cite{Zecca5}. The SEDS approach is based on electromagnetic interactions \cite{cavalleri1}. Our results show that there is a clear difference between the predictions of QM and those of SEDS for a macroscopic quantum system coupled to the environmental particles which depends on the quantum behaviour of the macro-system.

The paper is organized as follows. In section 2, we describe principles of the standard form of the Schr\"{o}dinger equation which is called dimensionless analysis. In section 3, bilinear harmonic model of the environment is introduced. Then, the ground state is exactly evaluated for the system coupled to two particles of the environment. In section 4, the two- dimensional double-slit diffraction pattern is analyzed. Two limiting cases of the problem are illustrated and discussed afterwards, to show how the emergence of classicality in the system leads to the appearance of macro-type fringes in the interference patterns. In section 5, we analyze the double-slit interference pattern for macro-system based on SEDS and compare the results with QM. Finally, the results are discussed in the conclusion part.

\section{Dimensionless form of the Schr\"{o}dinger equation}

First, we introduce dimensionless parameters for an arbitrary quantum system. We define $R_0$ and $U_0$ as constant units of length and energy, respectively. Subsequently, for a particle of mass $M$, one can define the characteristic time $\tau_0$  as \cite{Taka}:
\begin{equation}
\label{eq1}
\tau_0=\frac{R_0}{(U_0/ M)^\frac{1}{2}}
\end{equation}
in which $U_0$ is in order of the kinetic energy of the system and the unit of momentum could be defined as  $P_0=(U_0 M)^\frac{1}{2}$. Also, the conjugate variables $q$ and $p$, as well as the time $t$ in the dimensionless forms are defined as:
\begin{eqnarray}
\label{eq2}
q=\frac{R}{R_0}, ~~~       p=\frac{P}{P_0},  ~~~           t=\frac{T}{\tau_0}
\end{eqnarray}

\noindent where $R$, $P$ and $T$ are the conventional position, momentum and time, respectively. Then, the following relations for the potential energy $V$ and the Hamiltonian $H_s$ could be introduced in the dimensionless regime as:
\begin{eqnarray}
\label{eq3}
V(\hat q)=\frac{U(\hat R)}{U_0},~~~     {\hat{H}_s}=\frac{\hat{H}_S}{U_0}
\end{eqnarray}
\noindent where $U(\hat R)$ and $\hat{H}_S$ are the potential energy and the Hamiltonian in the ordinary Schr\"{o}dinger equation. Finally, the dimensionless form of the Schr\"{o}dinger equation can be written as:
\begin{equation}
\label{eq4}
i\bar h\frac{d\psi(t)}{dt}={\hat{H}_s}\psi(t)
\end{equation}
\noindent where
\begin{equation}
\label{eq5}
{\hat{H}_s}=\frac{\hat{p}^2}{2}+V(\hat{q})
\end{equation}
\noindent and
\begin{eqnarray}
\label{eq6}
 [\hat{q},\hat{p}]=i\bar h
 \end{eqnarray}
Here, we define
 \begin{equation}
 \label{eq7}
 \bar h=\frac{\hbar}{U_0\tau_0}=\frac{\hbar}{{P_0}{R_0}}=\lbrace\frac{\hbar^2}{MU_0 R^2_0}\rbrace^\frac{1}{2}
 \end{equation}

In equations (6) and (7), instead of Planck constant $\hbar$, a new dimensionless parameter $\bar h$ appears, measured in units of the action $U_0 \tau_0$, on which the quantum nature of the system depends. The situation where $\bar h\ll1$ is called the quasi-classical situation. The values of $\bar h$ between 0.01 to 0.1 could show the macroscopic trait of the proposed system \cite{Taka}. In many applications of double-well potentials, $R_0$ is defined as characteristic length of resonance between left and right counterparts of the potential $U(\hat R)$ . So, $\bar h$ in (7) can be rewritten as:
\begin{equation}
\label{eq8}
\bar h=\lambdabar_0/R_0
\end{equation}
\noindent where $\lambdabar_0=\lambda_0 / 2\pi$. Here, $\lambda_0$ denotes the de Broglie wavelength of the central system. For a macroscopic quantum system , $\lambda_0$ is too small compared to $R_0$ which the latter is nearly a fixed value for known models of potential. Thus, regarding the quasi-classical systems, the condition $\bar h<0.1$ seems suitable for our future purposes. For smaller values of $\bar h$, the macro-system shows more classical trait.

One can also define $\omega_0=\tau_0^{-1}= {P_0}/ {R_0 M}$ in relation with the $\it{particle}$ aspect of a system in a resonance situation. Accordingly, for the $\it{wave}$ aspect of the system, one can consider another unit of momentum $P_0'$ resulted from the phase velocity $v_0'=\omega_0/ k_0 (k_0=2\pi / \lambda_0)$, so that, $P_0'=M\lambdabar_0\omega_0$. Then, using the relation (8), we conclude that:
\begin{equation}
\label{9}
\bar h=\frac{P_0'}{P_0}
\end{equation}
This is another demonstration of how the value of $\bar h$ can display the classicality of the system. For a macroscopic quantum system, the wave character is weakened, so that $P_0' \ll P_0$.
\section{Bilinear-Harmonic model of the environment}

Suppose that the central system is a quantum harmonic oscillator. The entire system is composed of the system and  two environmental  micro-oscillators. The total Hamiltonian could be written as :
 \begin{equation}
\label{eq10}
\hat{H}=\hat{H}_s+\hat{H}_e+\hat{H}_{se}
\end{equation}
\noindent where $\hat{H}_s$ is the Hamiltonian of the system and $\hat{H}_e$ and $\hat{H}_{se}$ are the environment and the interaction Hamiltonians, respectively.

In the dimensionless form, the following classes of Hamiltonian can be used \cite{Taka}:
\begin{equation}
\label{eq11}
{\hat{H}_s}=\frac{\hat{p}^2}{2}+V(\hat{q})
\end{equation}
\begin{equation}
\label{eq12}
\hat{H}_e={\sum_\alpha}[ \frac{\hat{p}^2_\alpha}{2}+\frac{\omega^2_\alpha}{2}{\hat x^2_\alpha-}\frac{\bar h\omega_\alpha}{2}]
\end{equation}
\begin{equation}
\label{eq13}
\hat H_{se}=-\sum_\alpha{\omega}^2_\alpha f_\alpha(\hat q)\hat x_\alpha+\frac{1}{2}\sum_\alpha \omega^2_\alpha \lbrace f_\alpha (\hat q)\rbrace ^2
\end{equation}
\noindent where ${\alpha}$ varies from 1 to 2, which ${\alpha}$ is the number of the environmental oscillators. In these relations, $ x_\alpha $, $ p_\alpha $ and $ \omega_\alpha $ denote the position, the momentum and the frequency of the  environmental particles, respectively. The last constant term in (12), which merely displaces the origin of energy, has been added for later convenience. Hereafter, for simplicity, we assume that $f_\alpha(q)={\gamma_\alpha}q$, where $ 0<\gamma_\alpha<1$ denotes the strength of the interaction between the system and the environment. So, one can reach the conclusion that $\hat H_{se}$ is linear both to $ \hat x_\alpha$ and $\hat q$. So, the model is called $\it{bilinear}$.

Considering the bilinear condition, the total form of Hamiltonian can be presented as:
\begin{widetext}
\begin{equation}
\label{eq14}
H=\frac{p^2}{2}+V_1(q)+\sum_\alpha[\frac{p^2_\alpha}{2}+\frac{\omega^2_\alpha}{2}x^2_\alpha(1-\gamma_\alpha)-\frac{\bar h\omega_\alpha}{2}]+\frac{1}{2}\sum_\alpha \gamma_\alpha\omega^2_\alpha(x_\alpha-q)^2
\end{equation}
\end{widetext}
where
\begin{equation}
\label{eq15}
V_1(q)=V(q)-\frac{1}{2}\lbrace \sum_\alpha{\gamma_\alpha}(1-\gamma_\alpha)\omega^2_\alpha\rbrace q^2
\end{equation}
\noindent Here, $V(q)=\frac{1}{2}\omega_e ^2 q^2$ is the potential and $ \omega_e $ is the vibration frequency of the system. This shows that the system feels the efficient potential $V_1(q)$. Moreover, each environmental oscillator $\alpha$ with spring constant  $(1-\gamma_\alpha)\omega^2_\alpha$ is coupled to the system with spring constant $\gamma_\alpha\omega^2_\alpha$.

Now, we obtain the wave function  of the central system in the ground state. The total Hamiltonian can be written as:
\begin{equation}
\label{eq16}
H=\frac{p^2}{2}+\frac{\omega^2 }{2}q^2+\sum_\alpha[ \frac{p^2_\alpha}{2}+\frac{\omega^2_\alpha}{2}x^2_\alpha-\omega^2_\alpha\gamma_\alpha q x_\alpha]
\end{equation}
where
\begin{equation}
\label{eq17}
\omega=[\omega_e ^2+\sum_\alpha \omega^2_\alpha \gamma^2_\alpha]^\frac{1}{2}     
\end{equation}

For decoupling the Hamiltonian, we define plus-minus position and momentum coordinates, $(x_+,x_-)$ and
 $(p_+, p_-)$, respectively, by  rotating of the position coordinates $(q',x_\alpha)$ and the momentums $(p',p_\alpha)$ as the following \cite{McDermott}:
\begin{equation}
\label{eq18}
\begin{pmatrix}
x_+  \\
x_-  
\end{pmatrix}
\begin{matrix}\\\mbox{}\end{matrix}
=\begin{pmatrix} cos\theta & sin\theta \\ -sin\theta & cos\theta \end{pmatrix}
\begin{pmatrix} q' \\
x_\alpha
\end{pmatrix} 
\end{equation}

\begin{equation}
\label{eq19}
\begin{pmatrix}

p_+  \\
p_-  
\end{pmatrix}
\begin{matrix}\\\mbox{}\end{matrix}
=\begin{pmatrix} cos\theta & sin\theta \\ -sin\theta & cos\theta \end{pmatrix}
\begin{pmatrix} {p'} \\
{p}_\alpha
\end{pmatrix}
\end{equation}

\noindent where
\begin{equation}
\label{eq20}
p'=\frac{p}{N} ,  ~~~  q'=\frac{q}{N}    
\end{equation}
and $N$ is the total number of the environmental oscillators, which in our study $N=2$. Now, we define:
\begin{equation}
\label{eq21}
\theta=\frac{1}{2}arctan[ \frac{\omega'^2_\alpha}{\omega^2 -\omega^2_\alpha}]
\end{equation}
\noindent where $\omega'_\alpha=i ( 2{\omega^2_\alpha}{\gamma_\alpha})^\frac{1}{2}$. According to above decoupling method, the total Hamiltonian can be written as the sum of the individual Hamiltonians indexing $\alpha$:
\begin{equation}
\label{eq22}
{H}={\sum_\alpha}{H_\alpha}
\end{equation}
\noindent where
\begin{equation}
\label{eq23}
H_\alpha=\frac{p'^2}{2}+\frac{\omega^2}{2} q'^2+ \frac{p^2_\alpha}{2}+\frac{\omega^2_\alpha}{2}x
^2_\alpha-\omega^2_\alpha \gamma_\alpha q' x_\alpha
\end{equation}

Under the rotation, the kinetic energy part in Hamiltonian (23) remaines invariant. Thus, decoupling of the Hamiltonian is obtained by diagonalizing the potential energy. The rotations transform the Hamiltonian to 
\begin{equation}
\label{eq24}
H_\alpha=\frac{p_{+\alpha}^2}{2}+\frac{1}{2} \omega_{+\alpha}^2 x_{+\alpha}^2+\frac{p_{-\alpha}^2}{2}+\frac{1}{2} \omega^2_{-\alpha} x_{-\alpha}^2
\end{equation}
\noindent where $H_\alpha =H_{+\alpha}+H_{-\alpha}$.

\noindent Here, we define:
\begin{equation}
\label{eq25}
\omega_{+\alpha}=\lbrace \omega^2 \cos^2\theta+\omega^2_\alpha \sin^2\theta+\omega'^2_\alpha \sin\theta\cos\theta\rbrace^\frac{1}{2}
\end{equation}
and
\begin{equation}
\label{eq26}
\omega_{-\alpha}=\lbrace \omega^2 \sin^2\theta+\omega^2_\alpha \cos^2\theta-\omega'^2_\alpha \sin\theta\cos\theta\rbrace^\frac{1}{2}
\end{equation}

We also assume that for two particles of the environment, $\omega_\alpha$ and $\gamma_\alpha$ are nearly the same. Then, if $tan2\theta>0$ in (21), we should have $\omega^2<\omega^2_\alpha$. This means that
\begin{equation}
\label{eq27}
\omega^2_e<\omega^2_\alpha(1-\gamma^2 N)
\end{equation}

\noindent where $\omega^2_e=\omega^2-\omega^2_\alpha\gamma^2 N$ (see(17)) and $\gamma_\alpha=\gamma$. Yet, in (27), it is necessary that $(1-\gamma^2 N)>0$, or $N\gamma^2 <1$. In other words, the number of particles in the environment should restrict the strength of interaction $\gamma$, which is not a legitimate condition. On the other hand, if we take $tan2\theta<0$, it will be obtained from (9), (17) and (21) that
\begin{equation}
\label{eq28}
\lambda^2_0(1-\gamma^2 N)<\lambda^2_\alpha
\end{equation}

\noindent where $\lambda_\alpha$ is the wavelength of the environmental particles. For both small values of $N$ and $\gamma$ (so that $1-\gamma^2 N\approx1$), one concludes from (28) that
\begin{equation}
\label{eq29}
\bar h=\lambdabar_0/R_0<\lambdabar_\alpha/R_0
\end{equation}
\noindent which guarantees the quasi-classical behavior of the central system. Because, the characteristic wavelength of the macro-system $\lambda_0$ is much smaller than the corresponding wavelength $\lambda_\alpha$ of the micro-particles of the environment. We choose $0.01<\bar h=\lambdabar_0/R_0<0.1$ in (8) to reach the definite bound of $\bar h$ for a quasi-classical system, as mentioned before. Also, in (28), one can notice that $N\gamma^2 >(1-\lambda^2_\alpha / \lambda^2_0)$. If, one assumes that $\lambda_\alpha / \lambda_0<1$ (contrary to (29)), the values of $N$ and $\gamma$ will be again restricted to a positive constant value which is not reasonable, since they are independent parameters. So, for any value of $N$ and $0<\gamma<1$, $\lambda_\alpha / \lambda_0>1$, the condition of (29) is compelling. In effect, the situation for having a macroscopic quantum system is now ready. The key point is that the emergence of classicality, here, is due to the conditions the system interacts under which with the environment.

For calculating  the wave function of the system, we define $a=\sin\theta$ and $b=\cos\theta$. Using these definitions, the plus-minus position coordinates of two particles of the environment can be presented as:
\begin{align}
\label{eq30}
x_{+i}&=bq'+a{x}_i \nonumber\\
x_{-i}&=-aq'+b{x}_i
\end{align}

\noindent where $i=1,2$. In this case, the normal ground state wave function for Hamiltonian (24) can be obtained as:
\begin{widetext}
\begin{equation}
\label{eq31}
\psi_0(x_{+1} , x_{-1},x_{+2},x_{-2})=(\frac{{\omega_{+1}} {\omega_{-1}}{\omega_{+2}}{\omega_{-2}}}{{\pi^4 \bar h^4}})^\frac{1}{4}\exp(\frac{-{\omega_{+1}}{x_{+1}^2}}{2\bar h})\exp(\frac{-{\omega_{-1}}{x_{-1}^2}}{2\bar h})\exp(\frac{-{\omega_{+2}}{x_{+2}^2}}{2\bar h})\exp(\frac{-{\omega_{-2}}{x_{-2}^2}}{2\bar h})
\end{equation}
\end{widetext}
Then, one gets the probability distribution for the position coordinate of the system, $P(q)$, by integrating  the probability density over the spatial coordinates of the environmental oscillators ($x_{1} , x_{2}$).

\noindent  Considering the relations (30) and (31), we obtain:
\begin{widetext}
\begin{align}
\label{eq32}
P(q)=&\frac{1}{\pi \bar h}(\omega_{+1} \omega_{-1}\omega_{+2}\omega_{-2})^\frac{1}{2}[(a^2\omega_{+1}+b^2\omega_{-1})(a^2\omega_{+2}+b^2\omega_{-2})]^\frac{-1}{2} \nonumber\\
&\times\exp\lbrace {\frac{-q^2 [(a^2\omega_{+1}\omega_{+2})(\omega_{-1}+\omega_{-2})+(b^2\omega_{-1}\omega_{-2})(\omega_{+1}+\omega_{+2})]}{\bar h(a^2\omega_{+1}+b^2\omega_{-1})(a^2\omega_{+2}+b^2\omega_{-2})}\rbrace} 
\end{align}
\end{widetext}

Using the probability distribution (32), after some mathematical manipulation, one gets  the normal wave function for the  system as the following:
\begin{widetext}
\begin{align}
\label{eq33}
\psi(q)=&(\frac{1}{\pi \bar h})^\frac{1}{4}[(a^2\omega_{+1}+b^2\omega_{-1})(a^2\omega_{+2}+b^2\omega_{-2})]^\frac{-1}{4}
 {[(a^2{\omega_{+1}}{\omega_{+2}})(\omega_{-1}+\omega_{-2})+(b^2{\omega_{-1}}{\omega_{-2}})(\omega_{+1}+\omega_{+2})]}^\frac{1}{4}\nonumber\\
&\times\exp\lbrace {\frac{-q^2 [(a^2\omega_{+1}\omega_{+2})(\omega_{-1}+\omega_{-2})+(b^2\omega_{-1}\omega_{-2})(\omega_{+1}+\omega_{+2})]}{2\bar h(a^2\omega_{+1}+b^2\omega_{-1})(a^2\omega_{+2}+b^2\omega_{-2})}\rbrace} 
\end{align}
\end{widetext}

In the next section, we will formulate the double-slit diffraction pattern, using the wave function of the system in (33).

\section{Double-slit diffraction pattern based on QM approach}
Let us consider the system described in the previous section in two dimensions. We assume that the system has been in interaction with the environment only in $y$-direction, so that regarding the $x$-direction, the state of the system behaves like a Gaussian wave packet independent of any environmental effect. After the slits in both directions $x$ and $y$, the system can be viewed as a free particle, for which the wave function in the $y$-direction is defined as (33) at $t=0$ (just after the slits) where $q\equiv y$. The system has a macroscopic quantum character due to the conditions elaborated in section 3 (see relation (29)). Now, we analyze the two-slit diffraction pattern in two dimensions. For this purpose, first we define  the region $R$ that is inaccessible to the particle, assumed to be a subset of the $(x, y)$ plane:
\begin{equation}
\label{eq34}
R= \lbrace(x, y): \mid x \mid<a, y\in (-\infty, -d'-b']\cup [-d', d]\cup[d+b, \infty)\rbrace
\end{equation}

\noindent where $b$ and $b'$ are the widths of the slits with depth $2a$ ($b,b'\ll 1$) and $d$ and $d'$ are the distances of the slits from the origin \cite{Zecca4}.

The Hamiltonian of the system at $t\geq 0$ is defined as:
\begin{equation}
\label{eq35}
H=H_0+V(x,y)
\end{equation}
where $V(x,y)$ is an infinite step potential which does not allow the particle to tunnel to the region $R$. For solving the Schr\"{o}dinger equation, which in general could not be separated in terms of $x$ and $y$ dependences, one can use factorized solutions as legitimate method. The diffraction pattern of a particle which is defined by (33), can be studied in the following way. We consider a Gaussian wave packet coming from the remote $x$ region with probability distribution centered on a point moving with velocity $\bar h k_{0x}$ on the $x$-axis ($y=0$) \cite{Zecca4, Zecca5}:
\begin{equation}
\label{eq36}
\psi(x, y, t)=\chi(x, t)\phi(y,t)
\end{equation}
\noindent where
\begin{widetext}
\begin{equation}
\label{eq37}
\chi(x, t)=[\frac{\zeta}{\pi^\frac{1}{2}(1+i\bar h\zeta^2t)}]^\frac{1}{2}\lbrace exp[-\frac{\zeta^2}{2}\frac{(x-x_0-k_{0x}t )^2}{1+i\bar h\zeta^2 t}+\frac{ik_{0x}}{\bar h}(x-x_0)-ik_{0x}^2t/2\bar h]\rbrace
\end{equation}
\end{widetext}

Here, $\zeta$ is of the order of $(\omega_e/ \bar h)^\frac{1}{2}$ and $k_{0x}=\lambdabar_0/\lambdabar$ where $\lambda$ is the wavelength of the wave packet in the $x$-direction $(\lambdabar=\lambda/ 2\pi)$. We have also:
\begin{equation}
\label{eq38}
\phi(y, t)=[\frac{\beta}{\pi^\frac{1}{2}(1+i\bar h\beta^2t)}]^\frac{1}{2}exp[-\frac{\beta^2}{2}\frac{(y-y_0)^2}{1+i\bar h\beta^2 t}]
\end{equation}

\noindent where for the wave function of the central system in (33) at $t=0$, $\beta$ can be defined as:
\begin{widetext}
\begin{equation}
\label{eq39}
\beta=\lbrace\frac{[(a^2{\omega_{+1}}{\omega_{+2}})(\omega_{-1}+\omega_{-2})+(b^2{\omega_{-1}}{\omega_{-2}})(\omega_{+1}+\omega_{+2})]}{\bar h(a^2\omega_{+1}+b^2\omega_{-1})(a^2\omega_{+2}+b^2\omega_{-2})}\rbrace ^\frac{1}{2}
\end{equation}
\end{widetext}

\noindent All parameters and variables in relations (37) to (39) are dimensionless. For evaluating the behavior of the wave packet after the slits, we assume that the part of the wave function $\psi(x, y, t)$ relative to the points $(x, y)$ (such that $\mid y \mid >b $ ) is reflected towards the negative $x$-regions by the barrier $R$, because no tunneling effect is possible with an infinite potential barrier. So, we assume that the wave packet after the slits, at initial time $t=0$, can be presented as:
\begin{equation}
\label{eq40}
\psi_I(x, y, 0)=\chi_a(x, 0)\phi_I(y, 0)
\end{equation}

\noindent where $\chi_a(x, 0)$ is $\chi(x, 0)$ in (37) with $x_0=a$ and $\phi_I(y, 0)$ is $\phi(y, 0)$ in (38), now defined in two-interval set $I=[d,d+b]\cup [-d'-b',-d']$. Since the initial wave function is separated by its $x$ and $y$ constituent functions and the particle freely moves after the slits, one can deduce that:
\begin{equation}
\label{eq41}
\psi_I(x, y, t)=\chi_a(x, t)\phi_I(y, t)
\end{equation}

\noindent where $\chi_a(x, t)$ is defined in (37) with $x_0=a$. On the other hand, after the slits, the wave function in the $y$-direction evolves as: 
\begin{widetext}
\begin{align}
\label{eq42}
\phi_I(y, t)=\frac{1}{2\bar h}\frac{\beta^\frac{1}{2}}{\pi^\frac{5}{4}}\int_ {-\infty}^{+\infty} exp[\frac{i}{\bar h}(p_yy-\frac{p^2_yt}{2})]dp_y \int_ {I} exp[-\frac{i}{\bar h}p_y\xi-\frac{\beta^2}{2}(\xi-y_0)^2]d\xi
\end{align}
\end{widetext}

\noindent where the first integral is the Fourier transform of time evolution of the wave function in the momentum space and the last integral is the wave function of the particle in the momentum space at $t=0$. 

Integrating over the variable $p_y$, the above wave function can be presented as:
\begin{widetext}
\begin{align}
\label{eq43}
\phi_I(y, t)=(\frac{\beta}{2\pi^\frac{3}{2}i\bar ht})^\frac{1}{2} exp[y^2\frac{i}{2\bar ht}-y_0^2\frac{\beta^2}{2}]\int_ {I} exp[-\xi^2(\frac{\beta^2}{2}-\frac{i}{2\bar ht})+\xi(y_0\beta^2-\frac{iy}{2\bar ht})]d\xi
\end{align}
\end{widetext}

In the next section, we will analyze two limiting cases of the diffraction pattern, resulting from the wave function in (43).

\subsection{Limiting cases of the problem}

In the following, we consider two different limiting situations:

1) We suppose that the wave packet reaching the slits is narrower than both the slits, i.e.,
\begin{equation}
\label{eq44}
\Delta y=\frac{1}{\beta\sqrt{2}}\ll b, b'
\end{equation}

\noindent The above condition shows that $\beta$ is very large and $b, b' \ll 1$. In this case, from (43) one can show that:
\begin{widetext}
\begin{align}
\label{eq45}
\phi_I(y, t) \phi_I^*(y, t)\cong  \frac{\pi^\frac{-3}{2}\beta}{(1+\bar h^2t^2\beta^4)^\frac{1}{2}} exp[{-\frac{\beta^2(y-y_0)^2}{1+\bar h^2t^2\beta^4}}]
[\int_ {(y_0+d')\beta /\sqrt{2}}^{(y_0+b'+d')\beta /\sqrt{2}}exp(-t^2)dt+\int_ {(y_0-d-b)\beta /\sqrt{2}}^{(y_0-d)\beta /\sqrt{2}}exp(-t^2)dt]^2
\end{align}
\end{widetext}

\noindent This shows that $\phi_I \phi_I^*$ is a Gaussian-like distribution \cite{Zecca4, Zecca5}. Thus, the situation described in this case corresponds to an incident wave packet which after the slits, essentially remains  undisturbed in its configuration or is reflected toward the negative $x$-direction, according to whether the incoming y-probability distribution is centered with regard to one of the slits or not. In this case, the diffraction pattern, for any value of $\bar h$ between  $0.01<\bar h<0.1$ has a Gaussian form. No interference pattern is seen here, since the particle passes through one slit. The result is shown in Fig.1. For simplicity we have assumed that $y_0=0$. 

\begin{figure}[H]
\centering
\includegraphics[scale=0.5]{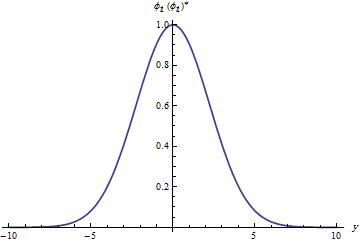}
\caption{Interference pattern for the case in which the incoming wave packet is too narrow with respect to both slits. }
\end{figure}

2) Now, we assume that the wave packet reaching the slits has very large uncertainty, depicted by the y-position probability distribution, i.e.,
\begin{equation}
\label{eq46}
\Delta y=\frac{1}{\beta\sqrt{2}}\gg b, b'
\end{equation}
\noindent From (46), it is evident that $\beta$ should be very small. By setting $\beta^2\approx 0$ in (43) and neglecting the term $i\xi^2 / 2\bar ht$ against $iy\xi / 2\bar ht$ for large values of $y$ (knows as far-field approximation), the integral term can be obtained as:
\begin{equation}
\label{eq47}
\frac{2\bar ht}{y}\lbrace exp[-\frac{iy}{\bar ht}(d+\frac{b}{2})] sin \frac{by}{2\bar ht}+exp[\frac{iy}{\bar ht}(d'+\frac{b'}{2})] sin \frac{b'y}{2\bar ht}\rbrace
\end{equation}
\noindent Assuming that $d=d'$, $b=b'$, from (43) and (47) one gets:
\begin{widetext}
\begin{equation}
\label{eq48}
\phi_I(y, t) \phi_I^*(y, t)\cong \frac{2b^2\beta}{\pi^\frac{3}{2}\bar ht} exp[-\beta^2 y_0^2] \frac{sin^2(by / 2\bar ht)}{(by/2\bar ht)^2} cos^2[\frac{y}{\bar ht} (d+\frac{b}{2})]
\end{equation}
\end{widetext}

As expected, this probability has a maximum at $y_0=0$. If  the separation of the slits is of the order of the slit width $(d\cong b)$, the factor containing the \textit{cosine} will be practically negligible and the expression (48) essentially gives the elementary diffraction pattern of a plane wave passing through a single slit. If the separation of the slits is much greater than their width $(d\gg b)$, the relation (48) represents a high-frequency pattern modulated by  elementary diffraction fringes. In this latter case, the resulting diffraction pattern is plotted in different situations with $\bar h=0.1$ and $\bar h=0.01$. The case for which $\bar h=0.1$ shows the quantum trait of the system and for $\bar h=0.01$, the pattern describes the classical behavior of the macro-system (see Fig.2). This is in agreement with what we expect for interference fringes of a macroscopic quantum system, when the value of $\bar h$ is sufficiently small ($\bar h < 0.1$) \cite{Zecca4, Zecca5}.

\begin{figure}
{\includegraphics[scale=0.5]{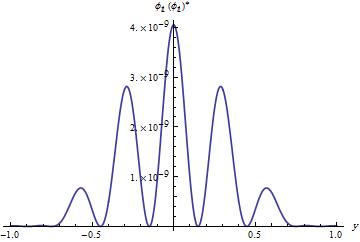}}
{\includegraphics[scale=0.5]{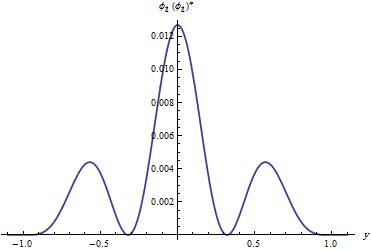}}
\caption{Interference pattern for the case in which the incoming wave packet has a great uncertainty $y$-direction with Left) $\bar h=0.1$ and Right) $\bar h=0.01$.}
\end{figure}

\section{Double-slit diffraction pattern based on SEDS approach}

In SEDS, the Schr\"{o}dinger equation is considered as a rough approximation for a stochastic process, which works well for the average of the trajectories of each state under some definite conditions (for example, for electrons bound in atoms), but not for single trajectories which occurs for the scattering process. A similar restriction holds for the diffraction of a beam of the particles passing through the two slits. The diffraction is due to the standing waves of the zero point field (ZPF) between the edges of the two slits that establishes itself across the clifts and depends on the classical spin motion with constant speed \cite{Zecca5, cavalleri2, cavalleri3}.

As a particle approaches one slit, its precession frequency increases and when it is equal to one of the standing ZPF waves between the two slits, it undergoes a transverse impulse from the ZPF with the following $y$-component of velocity:
\begin{equation}
\label{eq49}
v_y=\frac{\bar h\omega_n}{2c}
\end{equation}
where $c$ is the speed of light and $\omega_n$ is the angular velocity of the particle around the unit vector ($\hat n$) perpendicular to the plane of gyration orbit. The relation (49) is in dimensionless form. These random transversal impulses are maximum when $\omega_n$ coincides with a peak of the ZPF spectrum inside the slits. Notice that the transverse deviations should occur not only when the particle has equal probability to pass through either one or the other of the two slits, but also when it can pass through only one slit \cite{Zecca5}.

For the velocity vector, the deviation angle of the beam $\theta$ is defined as $sin\theta=\langle v_y^2\rangle^{1/2}/v$, where $v$ is the particle speed before and after crossing the slits. Using (49), one can obtain the following dimensionless relation
\begin{equation}
\label{eq50}
sin\theta=\pm \frac{\bar h\omega_n}{2cv}
\end{equation}
where $\bar h\omega_n$ is dimensionless energy per normal modes of the ZPF. The intensity of the deviated beam depends on the spatial density of modes allowed by the slits. The amplitudes of the ZPF waves are spatially uniform in space and zero on the wall of the slits. Consequently, the spatial Fourier transform of the ZPF amplitude is 
\begin{equation}
\label{eq51}
F_s=\frac{1}{2b}\int_ {-b}^{b} exp[ik_yy]dy=\frac{sin(k_yb)}{k_yb}
\end{equation}

The corresponding spatial distribution of the energy modes allowed by the slits is proportional to 
\begin{equation}
\label{eq52}
\rho_E \propto[sin(k_yb)/k_yb]^2
\end{equation}
which is familiar for ZPF waves and is also equivalent with what can be obtained from (48) for a wide beam (i.e., $\Delta y\gg b$) \cite{Zecca5}. The intensity maxima occur for 
\begin{equation}
\label{eq53}
k_y=0  \quad \textrm{and} \quad k_yb=\pi (n+\frac{1}{2}), \quad n=1,2,3,....
\end{equation}

Since $k_y=\omega/c$, one can show that this corresponds to 
\begin{equation}
\label{eq54}
sin\theta_m=0 \quad \textrm{and} \quad sin\theta_m=\pm \frac{\bar h\pi}{2bv}(n+\frac{1}{2})
\end{equation}

For a given $r$, where $r$ is the distance of the particle from the nearest edge of the slit and $\omega=v/r$, three maxima should appear: one for the central undeviated beam, and two others for two different points with opposite signs. This resembles the quantum prediction, specially when the open macro-system behaves more classically (see Fig.2 for $\bar h=0.01$). When the order of classicality is low (e.g., $\bar h=0.1$), the interference pattern has more details.

Yet, the interaction of the ZPF waves with the incident beam does not depend on the size of the slits as well as the width of the incident beam itself. So, the diffraction pattern should appear again, if the beam is much narrower than the slit width (i.e., $\Delta y\ll b$). This prediction is totally different with QM, because according to QM, no diffraction should occur in this condition (see the relation (45) and Fig.1). Therefore, the prediction of SEDS for a narrow beam of the particles with respect to the widths of the slits is that for each position of the entering beam, there are two angles of deviation given by (50). The average intensity of the two deviated beams is given by (52). Clearly, there should be always a central, non-deviated beam. By displacing gradually the position of the entering beam and registering the successive pairs of the opposite spots, the complete diffraction pattern can be observed for a wide beam (i.e., $\Delta y\gg b$), as mentioned before \cite{Zecca5,cavalleri1}.

We have drawn in Fig.3 the deviation angle $\theta$ against the parameter $\bar h$ and the widths of the slits $b$. As is clear, when the angle $\theta$ decreases, along with an increase in the widths of the slits $b$, the quantum nature of the macro-system is evanesced corresponding to smaller values of $\bar h$. In this situation, we expect that no interference occurs, but SEDS predict some fringes, as explained before. On the other hand, the same dependence of $\bar h$ could not be observed for the narrow slits, since the deviation angle $\theta$ increases. In this cases, SEDS predicts somehow an invariant interference pattern like for wide slits. Nevertheless, as we see in Fig.2, for an open quantum system, the interference fringes could be variant, depending on the nature of the system (illustrated by $\bar h$). This shows that there is a basic relationship between the quantum behaviour of the macro-system and the double-slit interference pattern predicted by SEDS.

\begin{figure}[H]
\centering
\includegraphics[scale=0.5]{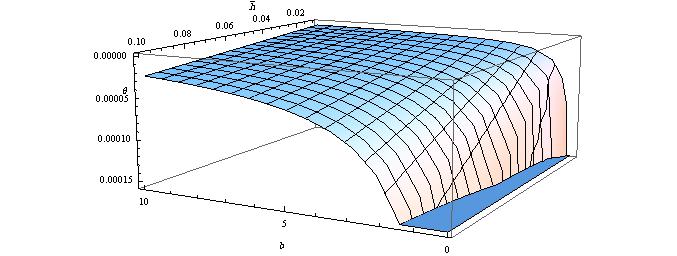}
\caption{The dependence of the deviation angle $\theta$ on the parameter $\bar h$ and the slits width $b$ for a macroscopic quantum system coupled to the environmental degrees of freedom. }
\end{figure}

Consequently, in the SEDS approach the interference fringes could be always present but with fixed patterns, since any particle passing through one of the two slits constantly feels the ZPF waves due to the boundary conditions. In other words, the spatial waves of the ZPF in (51) are always present between the two slits.

\section{Conclusion}

Interference patterns for different quantum systems have been considered for many decades. In recent years, however, the experts have encountered $\it{how}$ the quantum-to-classical transition occurrs when the system shows classical trait. Taking into account the effects of an interacting environment on a quantum harmonic system via a simple oscillating model, we have shown that when the quantumness of the system is evanesced (measured by the parameter $\bar h$ in (29)), interference fringes are diminished in accordance with known patterns observed for macro-molecules (see Fig. 2b) \cite{Nairz}. The environmental effects are not important when the incoming wave packet somehow describes the position state of the system in a given direction (see Fig.1). So, we have now a controllable parameter $\bar h$ by which we can follow and demonstrate the effects of the environment on quantum behavior of the system. This may open new door to the way one can better understand the emergence of classical appearance of the physical world in an interactive manner. 

Moreover, we compared the double-slit interference patterns obtained by QM and SEDS for an open macroscopic quantum system. Our results show that, contrary to what expected for closed systems, the diffraction pattern predicted by SEDS is not totally equal to the quantum case when the beam of particles is transversally uniform and much larger than the slit width $b$. In the latter case, the interference pattern shows different fringes for various value of $\bar h$. However, in the case of a beam that is narrow with respect to $b$, the difference is more apparent, because in this situation too, the SEDS theory predicts an interference pattern, while in quantum approach we see no interference. These results show that there is a clear difference between the predictions of QM and those of SEDS for an open macroscopic quantum system.

\end{document}